\documentclass[times,sort&compress,3p]{article}
\usepackage{geometry}
\newgeometry{vmargin={25mm}, hmargin={25mm,25mm}}
\usepackage[square,sort,comma,numbers]{natbib}

\usepackage{amsmath, amssymb, enumerate, amsthm}

\usepackage{multirow}		
\usepackage{comment}		
\usepackage{mathrsfs}		
\usepackage{tabularx,booktabs}		
\usepackage{siunitx}		

\usepackage{natbib}
\usepackage{chngcntr}
\usepackage{subcaption}
\usepackage{tikz}
\usepackage{float}
\usepackage{mathrsfs}
\usepackage{siunitx}
\usepackage{undertilde}
\usepackage[flushleft]{threeparttable}
\usepackage{booktabs}
\usepackage{multirow}
\usetikzlibrary{decorations.pathreplacing}
\usetikzlibrary{calc}
\usetikzlibrary{shapes}
\usepackage{scalerel,stackengine,amsmath}
\usepackage{lmodern}
\usepackage{xcolor,colortbl}

\tikzset{%
	add/.style args={#1 and #2}{
		to path={%
			($(\tikztostart)!-#1!(\tikztotarget)$)--($(\tikz\totarget)!-#2!(\tikztostart)$)%
			\tikztonodes},add/.default={.2 and .2}}
}

\colorlet{mfarbe}{red}
\colorlet{hidden}{white}
\definecolor{BlueLUH}{cmyk}{0.7,0.4,0,0}
\colorlet{LightBlue}{BlueLUH!20!white}

\definecolor{BlueMatlab}{rgb}{0,0.447,0.741}
\definecolor{OrangeMatlab}{rgb}{0.8500,0.3250,0.0980}
\definecolor{YellowMatlab}{rgb}{0.9290,0.6940,0.1250}

\newcommand\bone{3.5}
\newcommand\btwo{.1}
\newcommand\bthree{3}
\newcommand\bfour{3.5}
\newcommand\bfive{.8}

\usepackage{color}

\usepackage[
hidelinks
]{hyperref}		

\usepackage[figuresright]{rotating}

\newcommand{\N}{\mathbb{N}}
\newcommand{\R}{\mathbb{R}}

\newcommand{\hsigma}{\widehat{\sigma}}
\newcommand{\bfbeta}{\mathbf{f}_{\bbeta}}
\newcommand{\bFbeta}{\mathbf{F}_{\bbeta}}

\newcommand{\btX}{\boldsymbol{\widetilde{X}}}

\newcommand{\yinit}{y_{\text{init}}}
\newcommand{\hx}{\hat{x}}
\newcommand{\hy}{\hat{y}}

\newcommand{\No}{\operatorname{N}}

\newcommand{\CIboot}{\mathcal{C}^{\text{boot}}}
\newcommand{\CIasym}{\mathcal{C}^{\text{asym}}}

\newcommand{\PIboot}{\mathcal{P}^{\text{boot}}}
\newcommand{\PIasym}{\mathcal{P}^{\text{asym}}}

\newcommand{\bx}{\boldsymbol{x}}

\newcommand{\by}{\boldsymbol{y}}

\newcommand{\bz}{\boldsymbol{z}}

\newcommand{\bbeta}{\boldsymbol{\beta}}
\newcommand{\hbbeta}{\widehat{\boldsymbol{\beta}}}
\newcommand{\bgamma}{\boldsymbol{\gamma}}
\newcommand{\hbgamma}{\widehat{\boldsymbol{\gamma}}}
\newcommand{\btheta}{\boldsymbol{\theta}}
\newcommand{\hbtheta}{\widehat{\boldsymbol{\theta}}}

\newcommand{\Matlab}{\textsf{Matlab}\;}

\DeclareMathOperator*{\argmin}{arg\,min}

\newcommand{\otoprule}{\midrule[\heavyrulewidth]}

\graphicspath{{../../figures/}}

\begin{document}

\title{Modeling long-term capacity degradation of lithium-ion batteries}

\author{Marcus Johnen$^1$, Simon Pitzen$^1$, Udo Kamps$^1$, Maria Kateri$^1$\footnote{maria.kateri@rwth-aachen.de} \ and Dirk Uwe Sauer$^2$\\[1ex]
{\normalsize $^1$ Institute of Statistics, $^2$ Institute for Power Electronics and Electrical Drives (ISEA)}\\
        {\normalsize RWTH Aachen University, Germany}}

\date{}
\maketitle

\begin{abstract}
Capacity degradation of lithium-ion batteries under long-term cyclic aging is modelled via a flexible sigmoidal-type regression set-up, where the regression parameters can be interpreted. Different approaches known from the literature are discussed and compared with the new proposal. Statistical procedures, such as parameter estimation, confidence and prediction intervals are presented and applied to real data. The long-term capacity degradation model may be applied in second-life scenarios of batteries. Using some prior information or training data on the complete degradation path, the model can be fitted satisfactorily even if only short-term degradation data is available. The training data may arise from a single battery.\\[3ex]
\end{abstract}


\section{Introduction}\label{s:intro}

Lithium-ion batteries are used in a variety of systems, such as electric vehicles, grid storage applications, laptops and other electronic equipments. 
Thus, issues related to their safety and reliability are very crucial. 
A characteristic of major importance is the 
lifetime of batteries. Intensive research over the last years focuses on studying the aging mechanisms of batteries, targeting at predicting the remaining useful life of battery cells determined by their capacity over repetitive charge/discharge cycling.
The related bibliography is rich and the associated approaches are mainly classified in two categories: the model-based (e.g.\ \cite{Virkar:2011}) and the data-driven ones (e.g.\ \cite{Anton:2013,Ren:2018} and references cited therein), with the latter being mostly based on feature extraction and artificial intelligence methods. For a recent discussion of approaches in modeling lifetimes of batteries, we refer to \cite{wang:2019}.

Here, we focus on model-based approaches and, in particular, on a model for capacity degradation of batteries in a long-term study. In a data-driven approach, functional models for the batteries' performance are studied in order to gain some hints to physical processes and to provide a parametric model which may also be applied when dealing with second use of batteries (e.g., in a home storage system).  Once having accepted a long-term model, parameter fitting is thereafter possible with short-term experiments and/or by using small training samples. In the following, we propose and adopt a parametric model to serve as a
functional relationship between the cell capacity and cyclic aging of a lithium-ion battery. 
This model with five parameters is quite flexible and, as several examples show, can be fitted well to degradation data from long-term cyclic aging experiments.

The paper is organized as follows. In Section 2, the sigmoid parametric function is motivated and presented, emphasizing the role and physical interpretation of the parameters of the model. Parameter estimation and related algorithms as well as confidence and prediction intervals for the capacity at specific number of cycles are discussed in Section 3. Illustrative examples are presented in Section 4, focusing on early prediction of the cells' reliability and comparing the results to other standard models of the bibliography. The results are summarized in Section 5.

\section{A sigmoidal model proposal}\label{s:sigmoid}

\begin{figure}[ht]
\centering
\includegraphics[scale=0.3]{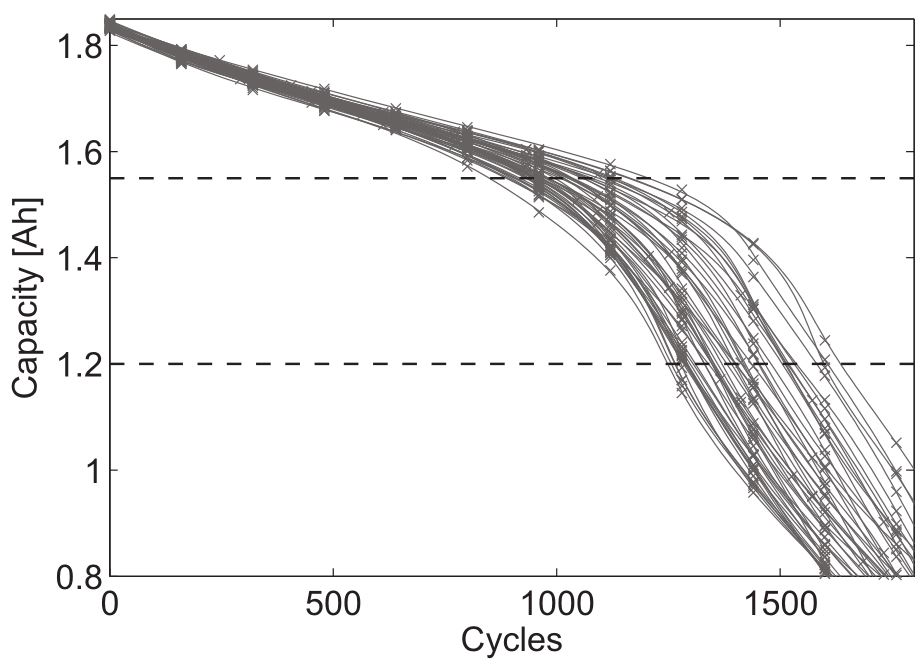}
\caption{Different aging trends from 48 cells from a single batch of cells under same aging conditions and profiles from \cite{baumhoefer:2014}.}
\label{fig:baumhoefer_cap_1}
\end{figure}

The capacity degradation of lithium-ion batteries, when observed until a low level of remaining capacity is reached, often shows a significant bend of the capacity curve, when a certain number of cycles is reached and the initial capacity has degraded to a certain amount. This can be seen, e.g., in \cite{baumhoefer:2014}, where 48 cells of the same type were aged under the same conditions (see Figure \ref{fig:baumhoefer_cap_1}). 
The degradation until about 80\% of the initial capacity is approximately linear. Thereafter, the degradation behavior changes.
For a detailed description of the experiment and the presented measurements we refer to \cite{baumhoefer:2014}.

\begin{figure}[ht]
\centering
\includegraphics[scale=1]{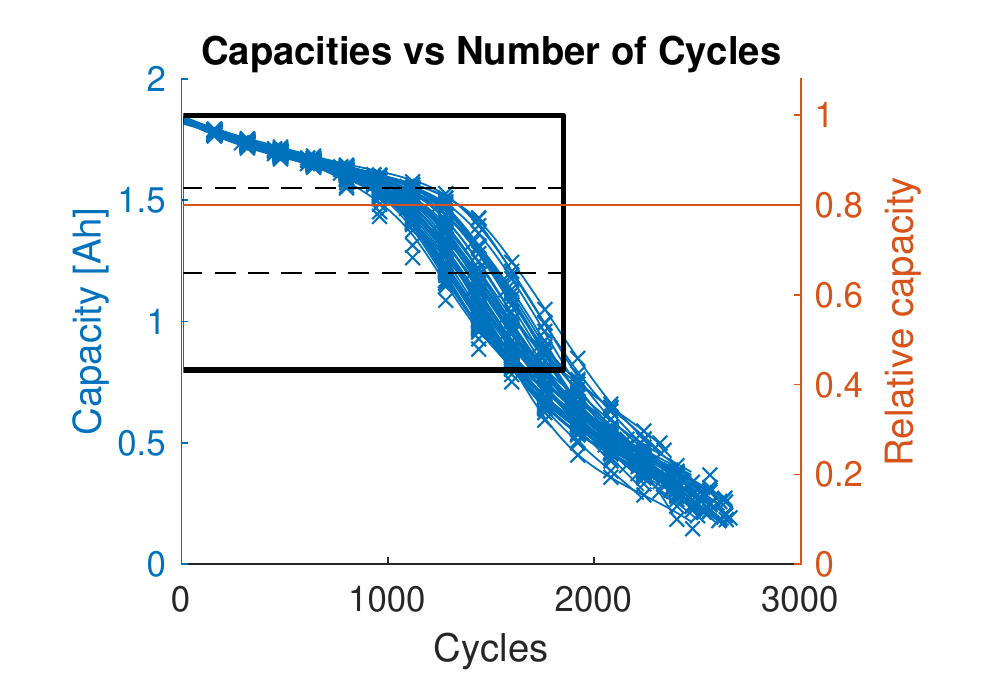}
\caption{Different aging trends from 48 cells from a single batch of cells under same aging conditions and profiles from \cite{baumhoefer:2014} measured longer than 2500 cycles. Results presented in \cite{baumhoefer:2014} (see Figure  \ref{fig:baumhoefer_cap_1}) are indicated by the window.}
\label{fig:baumhoefer_cap_2}
\end{figure}

However, this experiment by Baumh\"ofer et al.\ was continued beyond the results presented in \cite{baumhoefer:2014} and the additional data reveal a second bend at about 30\% of the initial capacity. Thus, as a whole, the capacity degradation follows an ``S"-shaped or sigmoidal curve as seen in Figure \ref{fig:baumhoefer_cap_2}, which is an extension of Figure \ref{fig:baumhoefer_cap_1} based on the complete sequence of data observed for the 48 cells in the experiment. Individual batteries from the two data sets considered in \cite{harrisharris:2017} and \cite{lewerenz:2017} exhibit a similiar behavior, from now on referred to as battery B and battery C, respectively. Their capacity degradation is shown in Figure \ref{fig:comp_model_fits}, along with that of a randomly chosen individual battery from \cite{baumhoefer:2014}, called battery A hereafter.

\begin{figure}[ht]
	\centering
	\includegraphics{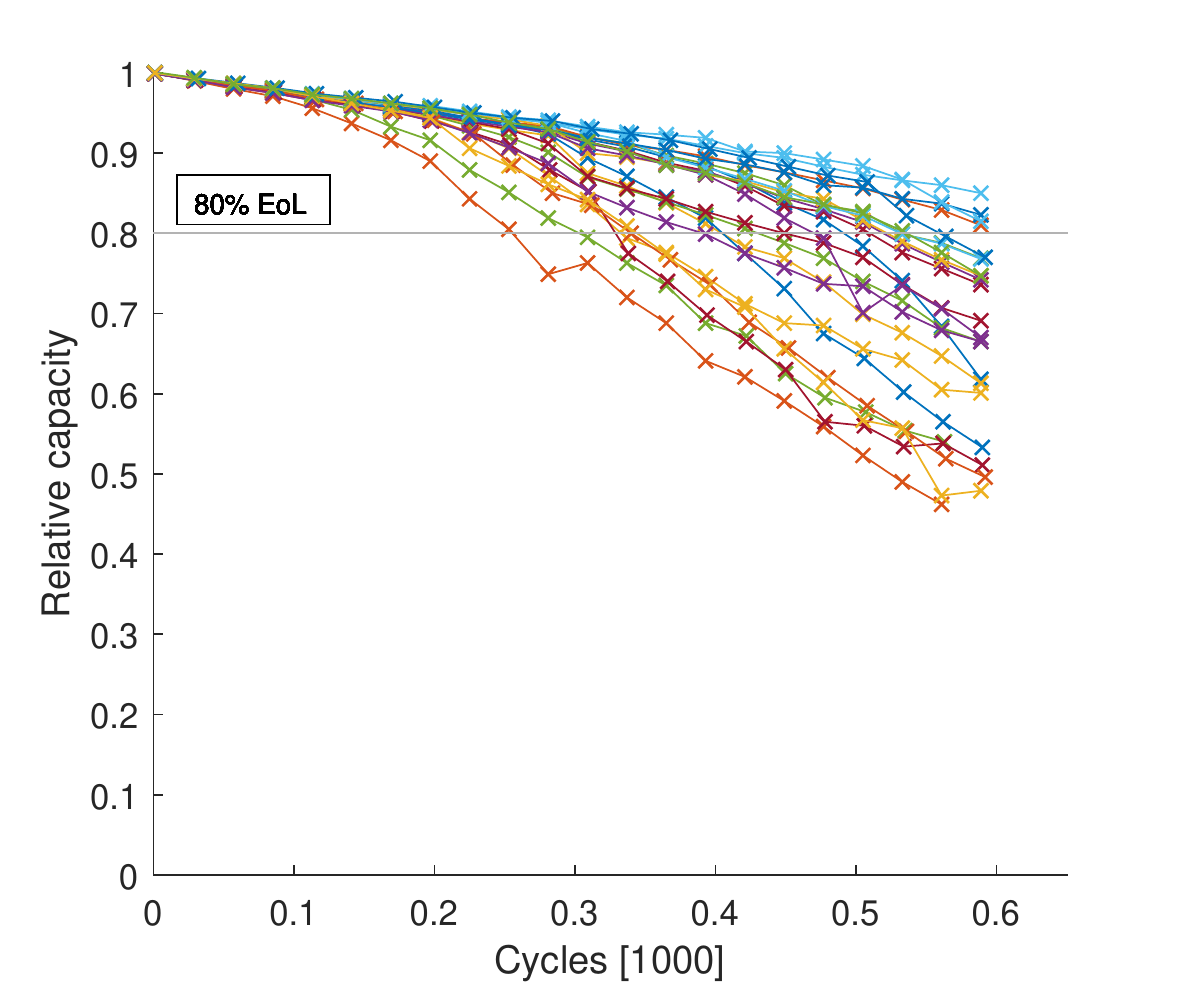}
	\caption{Aging trends from 24 cells from a single batch of cells under the same aging conditions as shown in \cite{harrisharris:2017}. Capacity information is available for every cycle, but only a fraction of those points is plotted in this figure to showcase the structure.}
	\label{fig:harris_daten}
\end{figure}

The functional connection of the measurement points in \cite{baumhoefer:2014} was obtained using a cubic spline interpolation, which is a convenient method to link the measured data via a smooth curve, capturing  the capacity degradation. For a statistical analysis of the degradation, we propose a parametric regression model which is then fitted to given battery data. By means of a stochastic model, the influence of measurement errors on the curve is reduced and statistical procedures can be applied.
Furthermore, our regression set-up is chosen in such a way that an interpretation of the parameters involved is possible.

In the literature, some parametric approaches to model the remaining capacity \(f(x;\boldsymbol{\beta})\) after \(x\) cycles depending on a parameter vector \(\boldsymbol{\beta}\) have been examined, such as a 

\textit{double exponential model} (cf.\ \cite{he:2011,crippspecht:2017, xing:2013, wang:2013, xu:2016})
\begin{align}
f(x;\boldsymbol{\beta}) = \beta_1e^{\beta_2x} + \beta_3e^{\beta_4x}, \quad x \ge 0,\; \boldsymbol{\beta} = (\beta_1, \beta_2, \beta_3, \beta_4)^\top \in \R^4, \label{eq:1}
\end{align}

a {\it polynomial model} (cf.\ \cite{xing:2013,micea:2011})
\begin{align}
f(x;\boldsymbol{\beta}) = \beta_1x^2 + \beta_2 x + \beta_3, \quad x \ge 0,\; \boldsymbol{\beta} = (\beta_1, \beta_2, \beta_3)^\top \in \R^3, \label{eq:2}
\end{align}

and a 
{\it mixture model} (cf.\ \cite{xing:2013, wang:2013})
\begin{align}
f(x;\boldsymbol{\beta}) = \beta_1 e^{\beta_2 x} + \beta_3 x^2 + \beta_4, \quad x \ge 0,\; \boldsymbol{\beta} = (\beta_1, \beta_2, \beta_3, \beta_4)^\top \in \R^4. \label{eq:3}
\end{align}
He et al.\ \cite{he:2011} apply the double exponential model (\ref{eq:1}) to describe the capacity degradation in reference to previous double exponential models for the increase of internal impedance which is strongly connected to the capacity fade of a battery. Monte Carlo methods are used to determine the model parameters and thus the remaining useful lifetime via extrapolation. Based on the work in \cite{he:2011}, Cripps and Pecht \cite{crippspecht:2017} develop a Bayesian, hierarchical nonlinear random effects model for longitudinal data to distinguish between two sets of batteries with differing quality.
Xu et al.\ \cite{xu:2016} introduce a Bayesian hierarchical model based on a spline approximation of the voltage profile to predict discharging profiles and the end of discharge. The results are used to estimate the remaining useful cycles as well, and the estimations show similar performance in comparison to the double exponential model.

Referring to \cite{bergveld:2002}, Micea et al.\ \cite{micea:2011} propose the polynomial model (\ref{eq:2}) for the capacity degradation of Nickel-Metal Hydride (Ni-MH) batteries. The model parameters are determined by a least squares approach to calculate the failure time for a single battery.

Motivated by the observation that the double exponential model and the polynomial model provide a better fit in respective data situations, Xing et al.\ \cite{xing:2013} present the mixture model (\ref{eq:3}) as the best choice from three different ensemble models with respect to the AIC. They follow a particle filter approach to predict the remaining useful performance by extrapolation of the fitted model. Wang et al.\ \cite{wang:2013} consider the double exponential model as justified in \cite{he:2011}, but observe that the parameters cannot be uniquely determined in a nonlinear least squares regression. Therefore, they replace one of the exponential terms by a power function with proposed exponent 2 or 4, depending on the observed degradation rate, and reduce the number of parameters to three. 
Moreover, we refer to Xu et al. \cite{xu:2018} for another parametric approach. \medskip\\


All models mentioned above perform well in the respective context and their respective range of application. However, none of them is flexible enough to yield a satisfying fit to capacity degradation data of a sigmoidal curve, which was observed in long-time experiments (cf.\ \cite{baumhoefer:2014}; see Figure \ref{fig:baumhoefer_cap_2}), as will be illustrated in Section \ref{s:examples}.

Motivated by this fact, we introduce a sigmoidal model for the capacity degradation using a combination of a linear trend and a logistic function:
\begin{align}\label{eq:4}
f(x;\boldsymbol{\beta}) = \beta_1 - \beta_2 x - \frac{\beta_3}{1+ \exp\left(-\frac{x-\beta_4}{\beta_5}\right)}, \quad x \ge 0,\; \boldsymbol{\beta} = (\beta_1,\dots,\beta_5)^\top \in (0,\infty)^5.
\end{align}
A shifted version of (\ref{eq:4}), given by 
\begin{align}
f(x;\boldsymbol{\beta}) = \beta_1 - \beta_2 x - \frac{\beta_3}{1+ \exp\left(-\frac{x-\beta_4}{\beta_5}\right)} + \frac{\beta_3}{1+ \exp\left(\frac{\beta_4}{\beta_5}\right)}, \quad x \ge 0,\; \boldsymbol{\beta} = (\beta_1,\dots,\beta_5)^\top \in (0,\infty)^5, \label{eq:5}
\end{align}
enables an interpretation of the parameter $\beta_1$ as the initial capacity $\yinit$ at $x=0$.

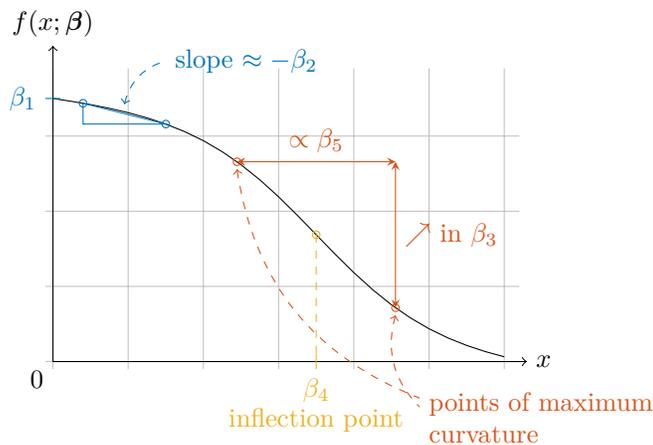
\begin{figure}[ht]
\centering
		\begin{tikzpicture}[domain=0:6]
		\draw[very thin,color=lightgray] (-0.1,-0.1) grid (6.2,3.9);
		\draw[->] (0,0) -- (6.3,0) node[right] {$x$};
		\draw[->] (0,0) node[below left] {0} -- (0,4.2) node[above] {$f(x;\bbeta)$};
		
		\draw[color=black] plot (\x,{\bone-\btwo*\x - \bthree/(1+exp(-(\x-\bfour)/\bfive)) + \bthree/(1+exp(\bfour/\bfive))});
		
		\draw[color=BlueMatlab] (.1,\bone) -- (-.1,\bone) node[left] {$\beta_1$};
		
		\draw[color=BlueMatlab] (0.4,{\bone-\btwo*0.4 - \bthree/(1+exp(-(0.4-\bfour)/\bfive)) + \bthree/(1+exp(\bfour/\bfive))}) node[circle,inner sep=0pt,minimum size=.1cm,draw] {}
		-- (0.4,{\bone-\btwo*1.5 - \bthree/(1+exp(-(1.5-\bfour)/\bfive)) + \bthree/(1+exp(\bfour/\bfive))})
		-- (1.5,{\bone-\btwo*1.5 - \bthree/(1+exp(-(1.5-\bfour)/\bfive)) + \bthree/(1+exp(\bfour/\bfive))}) node[circle,inner sep=0pt,minimum size=.1cm,draw] {}
		-- (0.4,{\bone-\btwo*0.4 - \bthree/(1+exp(-(0.4-\bfour)/\bfive)) + \bthree/(1+exp(\bfour/\bfive))});
		
		\draw[<-,color=BlueMatlab,dashed]  (0.95,{\bone-\btwo*0.95 - \bthree/(1+exp(-(0.95-\bfour)/\bfive)) + \bthree/(1+exp(\bfour/\bfive)) + 0.1}) to [bend left] (1.5,4) node[right] {slope $\approx -\beta_2$};
		
		\draw[color=YellowMatlab] (\bfour,{\bone-\btwo*\bfour - \bthree/(1+exp(-(\bfour-\bfour)/\bfive)) + \bthree/(1+exp(\bfour/\bfive))}) node[circle,inner sep=0pt,minimum size=.1cm,draw] {};
		\draw[dashed,color=YellowMatlab] (\bfour,{\bone-\btwo*\bfour - \bthree/(1+exp(-(\bfour-\bfour)/\bfive)) + \bthree/(1+exp(\bfour/\bfive))}) -- (\bfour,.1);
		\draw[color=YellowMatlab] (\bfour,.1) -- (\bfour,-.1) node[below] {$\beta_4$} (\bfour,-.5) node[below] {inflection point};
		
		\coordinate (x1) at ({\bfour-\bfive*ln(2-sqrt(3))},{\bone-\btwo*(\bfour-\bfive*ln(2-sqrt(3))) - \bthree/(1+exp(-((\bfour-\bfive*ln(2-sqrt(3)))-\bfour)/\bfive)) + \bthree/(1+exp(\bfour/\bfive))});
		\coordinate (x12) at ({\bfour-\bfive*ln(2-sqrt(3))},{\bone-\btwo*(\bfour-\bfive*ln(2+sqrt(3))) - \bthree/(1+exp(-((\bfour-\bfive*ln(2+sqrt(3)))-\bfour)/\bfive)) + \bthree/(1+exp(\bfour/\bfive))});
		\coordinate (x2) at ({\bfour-\bfive*ln(2+sqrt(3))},{\bone-\btwo*(\bfour-\bfive*ln(2+sqrt(3))) - \bthree/(1+exp(-((\bfour-\bfive*ln(2+sqrt(3)))-\bfour)/\bfive)) + \bthree/(1+exp(\bfour/\bfive))});
		
		\draw[color=OrangeMatlab] (x1) node[circle,inner sep=0pt,minimum size=.1cm,draw] {};
		\draw[color=OrangeMatlab] (6.5,-.75) node[text width=3cm](text) {points of maximum curvature};
		\draw[<-,dashed,color=OrangeMatlab] ($ (x2) + (0,-.1) $) to [bend right] (text);
		\draw[<-,dashed,color=OrangeMatlab] ($ (x1) + (0,-.1) $) to [bend right] (text);
		\draw[<->,>=stealth,color=OrangeMatlab] (x1) -- (x12) node [midway, right] {$\nearrow$ in $\beta_3$};
		\draw[<->,>=stealth,color=OrangeMatlab] (x12) -- (x2) node [midway, above] {$\propto \beta_5$} node[circle,inner sep=0pt,minimum size=.1cm,draw] {};

		\end{tikzpicture}
		\caption{Interpretation of the parameters $\beta_1,\ldots,\beta_5$ in the sigmoidal model.}
		\label{fig:param_sigmoid}
\end{figure}

The interpretation of all parameters of model (\ref{eq:5}) is illustrated in Figure \ref{fig:param_sigmoid}. 
Parameter  \(\beta_2\) describes approximately the negative slope in the linear section at the beginning of the curve, and \(\beta_4\) is the inflection point of the curve. The parameters \(\beta_3\) and \(\beta_5\) are related to the points of maximum curvature; their vertical distance is increasing in \(\beta_3\) and their horizontal distance is proportional to \(\beta_5\). Note that a similar sigmoidal approach has been proposed by \cite{Weng:2013} in the context of an open-circuit-voltage model for incremental capacity analysis of lithium-ion batteries.

With the logistic function $g$ given by $g(x)= g(x; \beta_3, \beta_4, \beta_5)=\beta_3/(1+ \exp(-(x-\beta_4)/\beta_5))$, we find 
\begin{equation*}
f(x;\boldsymbol{\beta}) = \beta_1 - \beta_2 x - g(x)+g(0) \ \ \text{and} \ \ f'(x;\boldsymbol{\beta}) = - \beta_2 - g'(x)
\end{equation*}
and $g$ satisfies the logistic or Verhulst differential equation
\begin{equation*}
g'(x) = \frac{1}{\beta_5}g(x)\left(1-\frac{g(x)}{\beta_3}\right)
\end{equation*}
which may give rise for a chemical or physical interpretation of the underlying battery degradation process.

\section{Model Estimation}\label{s:model_est}

Let $\by=(y_1,\dots,y_n)^\top\in\R^n$ denote the vector of capacity measurements after the cycles $\bx=(x_1,\dots,x_n)^\top\in\R^n$.
We next consider a stochastic model which incorporates the sigmoidal model as the expected capacity degradation of a battery. Measurement errors are then included by means of additive independent Gaussian random variables with constant variance. Formally, the resulting regression model can be stated as
\begin{align}\label{eq:regression_model}
y_j &= f(x_j;\bbeta) + \varepsilon_j, \quad j\in\{1,\dots,n\},
\end{align}
with $f$ as in \eqref{eq:5} and with measurement errors $\varepsilon_1,\dots,\varepsilon_n$. Hence, 
$\boldsymbol{\varepsilon}=(\varepsilon_1,\dots,\varepsilon_n)^\top\sim\No(\boldsymbol{0},\sigma^2\boldsymbol{I})$ for some constant but unknown variance $\sigma^2>0$, where \(\boldsymbol{0} = (0,\dots, 0)^\top \in \R^n\) and \(\boldsymbol{I} \in \R^{n \times n}\) denotes the \((n \times n)\)-identity matrix.

\subsection{Point estimation and lifetime prediction}\label{ss:estimation_prediction}
In the following, we discuss the estimation of the unknown parameter vector $\bbeta$. Estimation of $\bbeta$ is usually carried out by solving the corresponding (non-linear) least squares problem
\begin{equation}\label{eq:nls}
\hbbeta = \argmin_{\bbeta\in(0,\infty)^5}\sum_{j=1}^n\big(f(x_j; \bbeta)-y_j\big)^2.
\end{equation}
The resulting estimator $\hbbeta$ also corresponds to the maximum likelihood estimator in the stochastic model given in \eqref{eq:regression_model}. Technically, the minimization problem in \eqref{eq:nls} yields no closed-form solution for the parameter vector $\bbeta$ and iterative techniques are necessary. To solve \eqref{eq:nls} we first use the concept of conditional linearity, as described, e.g., in \cite[Ch.\ 3.5.5]{bateswatts:1988} and \cite[Ch.\ 14.7]{seberwild:2003}. This method, as described in the following, splits the solution of a non-linear least squares problem into a step-wise optimization on subvectors. Computationally, by reducing the dimension of the respective parameter spaces, the process converges faster and shows a lower risk of converging to a local minimum different from the optimal solution. 
 
In our particular case, we make use of the fact that $\bbeta$ can be split up into $\bgamma=(\beta_1,\beta_2,\beta_3)^\top$ and $\btheta=(\beta_4,\beta_5)^\top$, i.e., $\bbeta=(\bgamma,\btheta)$, such that $f$ from \eqref{eq:5} is linear in $\bgamma$ but not in $\btheta$. For this reason, for a fixed vector $\btheta$, the vector $\bgamma$ minimizing the sum of squares in \eqref{eq:nls} can be given in closed form as a function of $\btheta$ and it is therefore denoted by $\hbgamma(\btheta)$. In particular, from common theory on ordinary least squares, it follows that $\hbgamma(\btheta)$ can be computed as
\begin{align}\label{eq:nls_cond_lin2}
\hbgamma(\btheta) &= \left(\btX^\top\btX\right)^{-1}\btX^\top\by
\end{align}
where $\btX = \left( \boldsymbol{1} \mid -\bx \mid \bz(\bx,\btheta) \right)\in\R^{n\times 3}$ with $\boldsymbol{1}=(1,\dots,1)^\top\in\R^n$,  $\bz(\bx,\btheta)=(z_1,\dots,z_n)^\top\in\R^n$ and $z_j=1/(1+\exp(\theta_1/\theta_2))-1/(1+\exp(-(x_j-\theta_1)/\theta_2))$, $j\in\{1,\dots,n\}$. This reduces the situation in \eqref{eq:nls} to the optimization problem 
\begin{align}\label{eq:nls_cond_lin}
\hbtheta = \argmin_{\btheta\in(0,\infty)^2}\sum_{j=1}^n\big(f(x_j; (\hbgamma(\btheta),\btheta))-y_j\big)^2
\end{align}
involving only two parameters, which is then solved using a sequence of optimization algorithms, available within the software \Matlab which was used for all following calculations and simulations. More detailed, the \Matlab functions \textsf{ga}, \textsf{simulannealbnd}, and \textsf{lsqnonlin} are used sequentially to find the global minimum in \eqref{eq:nls_cond_lin}. Here, the resulting estimate of one algorithm is used as a starting value in the next algorithm. This procedure was chosen because, in general, the objective function to be minimized in \eqref{eq:nls_cond_lin} has multiple local minima corresponding to parameters $\btheta$ far from the global minimum. The resulting computational estimate $\hbbeta=(\hbgamma(\hbtheta),\hbtheta)$ is then an approximate solution to the original optimization problem \eqref{eq:nls}. Note that, for some \(\btheta \in \R^2\), the components of the solution \(\hbgamma(\btheta)\) of \eqref{eq:nls_cond_lin2} are not necessarily positive. However, if the components of \(\hbgamma(\hbtheta)\) are positive, the all-over solution \(\hbbeta=(\hbgamma(\hbtheta),\hbtheta)\) is still optimal for the constrained problem \eqref{eq:nls}. Otherwise, the problem \eqref{eq:nls} can be solved directly using the \Matlab functions \textsf{ga}, \textsf{simulannealbnd}, and \textsf{lsqnonlin} with bound constraints.

In practice, model \eqref{eq:regression_model} and the above estimation procedure can be applied to data from a single battery or from several batteries of the same type, targeting, in both cases, at lifetime prediction and derivation of confidence intervals. 
On the one hand, if the $n$ measurements in \eqref{eq:nls} belong to a single battery, the resulting estimate $\hbbeta$ describes the capacity degradation of that specific battery, only. On the other hand, if $y_1,\dots,y_n$ contain measurements from several batteries of the same type (e.g., all available measurements from an experiment), $\hbbeta$ describes the expected capacity degradation of a typical battery of this type. Both approaches are pursued in the examples of Section \ref{s:examples}. \bigskip

The fitted curve \(f(\,\cdot\,;\hbbeta)\) resulting from data of only one battery can be used for inter- and extrapolation of its remaining capacity after some cycle $x_0>0$ by $\hy = f(x_0;\hbbeta)$. The latter is not recommended due to instability outside of the range of measurements unless the available data exceed the inflection point illustrated in Figure \ref{fig:param_sigmoid}. If data from several batteries are used, the fitted curve can be used to predict the failure time \(x_q\), $q\in(0,1)$, of a new battery with respect to $q\cdot 100\%$ end of life (EoL), i.e.\ the time where the battery capacity falls below $y_q = q\cdot \yinit$, by \(\hx_q=f^{-1}(y_q;\hbbeta)\).

\subsection{Asymptotic confidence and prediction intervals}\label{ss:confidence_intervals_asymptotic}

The stochastic model from \eqref{eq:regression_model} allows for the construction of pointwise confidence and prediction intervals for the battery's expected capacity after some cycle. The interpretation of the two types of intervals depends on whether data from one or several batteries are used for inference, as discussed above in the model context. For a single battery, a confidence interval quantifies the uncertainty in estimating the capacity of that battery whereas for several batteries, it quantifies the uncertainty in estimating the mean capacity of the considered battery type. The corresponding prediction intervals account additionally for the measurement uncertainty.
First, approximate confidence and prediction intervals are derived which are based on asymptotic theory and are therefore valid for large sample sizes, only. For small samples, bootstrap confidence and prediction intervals should be preferred which are discussed in the next section.

The asymptotic confidence and prediction intervals for the battery capacity are derived by the method described in 
\cite[p.\ 193]{seberwild:2003}. 
Given data $(\bx,\by)$, for any cycle $x_0>0$, an approximate $(1-\alpha)$-level confidence interval $\CIasym_\alpha(x_0)$ for the expected capacity at cycle $x_0$, i.e. for $f(x_0;\bbeta)$, is given by
$\CIasym_\alpha(x_0)=[f(x_0;\hbbeta) - \delta, f(x_0;\hbbeta) + \delta]$, with 
\begin{equation*}
	\delta = t_{\frac{\alpha}{2}}(n-5)\,\hsigma\sqrt{\mathbf{f}_{\bbeta}(x_0;\hbbeta)^\top\big(\mathbf{F}_{\bbeta}(\bx;\hbbeta)^\top\,\mathbf{F}_{\bbeta}(\bx;\hbbeta)\big)^{-1}\, \mathbf{f}_{\bbeta}(x_0;\hbbeta)}\,
	\end{equation*}
where $t_{p}(d)$ denotes the $p$-quantile of the $t$-distribution with $d$ degrees of freedom,
$\bfbeta(x_0;\hbbeta)$ is the vector of partial derivatives of the sigmoid function with respect to $\bbeta$, evaluated at $\hbbeta$ for the cycle $x_0$, given by
	\begin{equation*}
	\bfbeta(x_0;\hbbeta) = \Bigg( \frac{\partial f}{\partial\beta_1}(x_0;\hbbeta)\,,\,\dots\,,\,\frac{\partial f}{\partial\beta_5}(x_0;\hbbeta) \Bigg)^\top\in\R^5,
	\end{equation*}
$\bFbeta(\bx;\hbbeta)$ is a related matrix of derivatives for the cycles $x_1,\dots,x_n$,
	\begin{equation*}
	\bFbeta(\bx;\hbbeta) = \Big(\bfbeta(x_1;\hbbeta) \,\big|\, \dots \,\big|\, \bfbeta(x_n;\hbbeta) \Big)^\top \in\R^{n \times 5},
	\end{equation*}
while $\hbbeta$ and $\hsigma^2$ are based on $(\bx,\by)$ and given by \eqref{eq:nls} and
	\begin{align}\label{eq:sigma_hat}
	\hsigma^2 = \frac{1}{n-5}\sum_{j=1}^n\big(f(x_j; \hbbeta)-y_j\big)^2 ,
	\end{align}
	respectively.

	A $(1-\alpha)$-level prediction interval $\PIasym_\alpha(x_0)$ for the capacity after cycle $x_0$, i.e. for $y_0=f(x_0;\bbeta)+\varepsilon$, $\varepsilon\sim\No(0,\sigma^2)$,  is given by $\PIasym_\alpha(x_0)=[f(x_0;\hbbeta) - \widetilde{\delta}, f(x_0;\hbbeta) + \widetilde{\delta}]$, where
		\begin{equation*}
		\widetilde{\delta} = t_{\frac{\alpha}{2}}(n-5)\,\hsigma\sqrt{1+\mathbf{f}_{\bbeta}(x_0;\hbbeta)^\top\big(\mathbf{F}_{\bbeta}(\bx;\hbbeta)^\top\,\mathbf{F}_{\bbeta}(\bx;\hbbeta)\big)^{-1}\, \mathbf{f}_{\bbeta}(x_0;\hbbeta)}\,.
		\end{equation*}

\subsection{Bootstrap confidence and prediction intervals}\label{ss:confidence_intervals_bootstrap}

Based on common bootstrap ideas (cf., e.g., Ch. 5 in \cite{efron:1982}), we propose a (parametric) bootstrap method, where the given measurement data are analyzed and the results are used to generate an independent random sample of error terms in order to calculate a bootstrap sample of capacity measurements based on the fitted sigmoidal curve many times. The resulting parameter estimations for the respective sigmoidal fits to the different samples are used to construct a two-sided bootstrap confidence interval of the sigmoidal fit at any arbitrary cycle.

Given data $(\bx,\by)$, for any $x_0>0$, an approximate $(1-\alpha)$-level bootstrap confidence interval $\CIboot_\alpha(x_0)$ for $f(x_0;\bbeta)$ using the bootstrap sample size $B\in\N$ is computed as follows:
\begin{enumerate}
	\item[(i)] Calculate $\hbbeta$ and $\hsigma^2$ based on $(\bx,\by)$ as in \eqref{eq:nls} and \eqref{eq:sigma_hat}, respectively.
	\item[(ii)] For $j\in\{1,\dots,n\}$ and $b\in\{1,\dots,B\}$, generate $\varepsilon_j^{(b)}$ from $\No(0,\widehat{\sigma}^2)$ independently and calculate $\by^{(b)}=(y_1^{(b)},\dots,y_n^{(b)})^\top$ with
	\begin{align*}
	y_j^{(b)} = f(x_j;\hbbeta) + \varepsilon_j^{(b)}.
	\end{align*}
	\item[(iii)] For each $b\in\{1,\dots,B\}$, derive the estimate $\hbbeta^{(b)}$ based on the sample $(\bx,\by^{(b)})$ as in \eqref{eq:nls}.
	\item[(iv)] Compute $\CIboot_\alpha(x_0)=[\ell,u]$, where $\ell$ and $u$ are the empirical $\frac{\alpha}{2}$- and $(1-\frac{\alpha}{2})$-quantile of the set of all $f(x_0;\hbbeta^{(b)})$, $b\in\{1,\dots,B\}$, respectively.
\end{enumerate}
	For an approximate $(1-\alpha)$-level prediction interval $\PIboot_\alpha(x_0)$ for the measured capacity $y_0=f(x_0;\bbeta)+\varepsilon$, $\varepsilon\sim\No(0,\sigma^2)$, after cycle $x_0$, step (iv) from above can be substituted by the following steps (iv') and (v'):
	\begin{itemize}
		\item[(iv')] For some $M\in\N$, generate independent and $\No(0,\hsigma^2)$-distributed $\varepsilon_m^{(b)}$, $m\in\{1,\dots,M\}$, $b\in\{1,\dots,B\}$, and calculate the resulting prediction errors $e_m^{(b)} = f(x_0;\hbbeta^{(b)}) - (f(x_0;\hbbeta) + \varepsilon_m^{(b)})$.
		\item[(v')] Compute $\PIboot_\alpha(x_0)=[f(x_0;\hbbeta^{(b)}) - \tilde{u},f(x_0;\hbbeta^{(b)}) - \tilde{\ell}]$, where $\tilde{\ell}$ and $\tilde{u}$ are the empirical $\frac{\alpha}{2}$- and $(1-\frac{\alpha}{2})$-quantile of the set of all $e_m^{(b)}$, $m\in\{1,\dots,M\}$, $b\in\{1,\dots,B\}$, respectively.
	\end{itemize}

The number of bootstrap samples \(B\) is chosen appropriately to the variation in the given situation and a value of $B=1000$ is used here. The procedures from Sections \ref{ss:confidence_intervals_asymptotic} and \ref{ss:confidence_intervals_bootstrap} are based on the validity of the model assumptions \eqref{eq:regression_model} and therefore may worsen if at least one of the assumptions is violated.

\section{Examples}\label{s:examples}

\subsection{Model comparison}

\begin{figure}[p]
	\centering
	\includegraphics[scale=1]{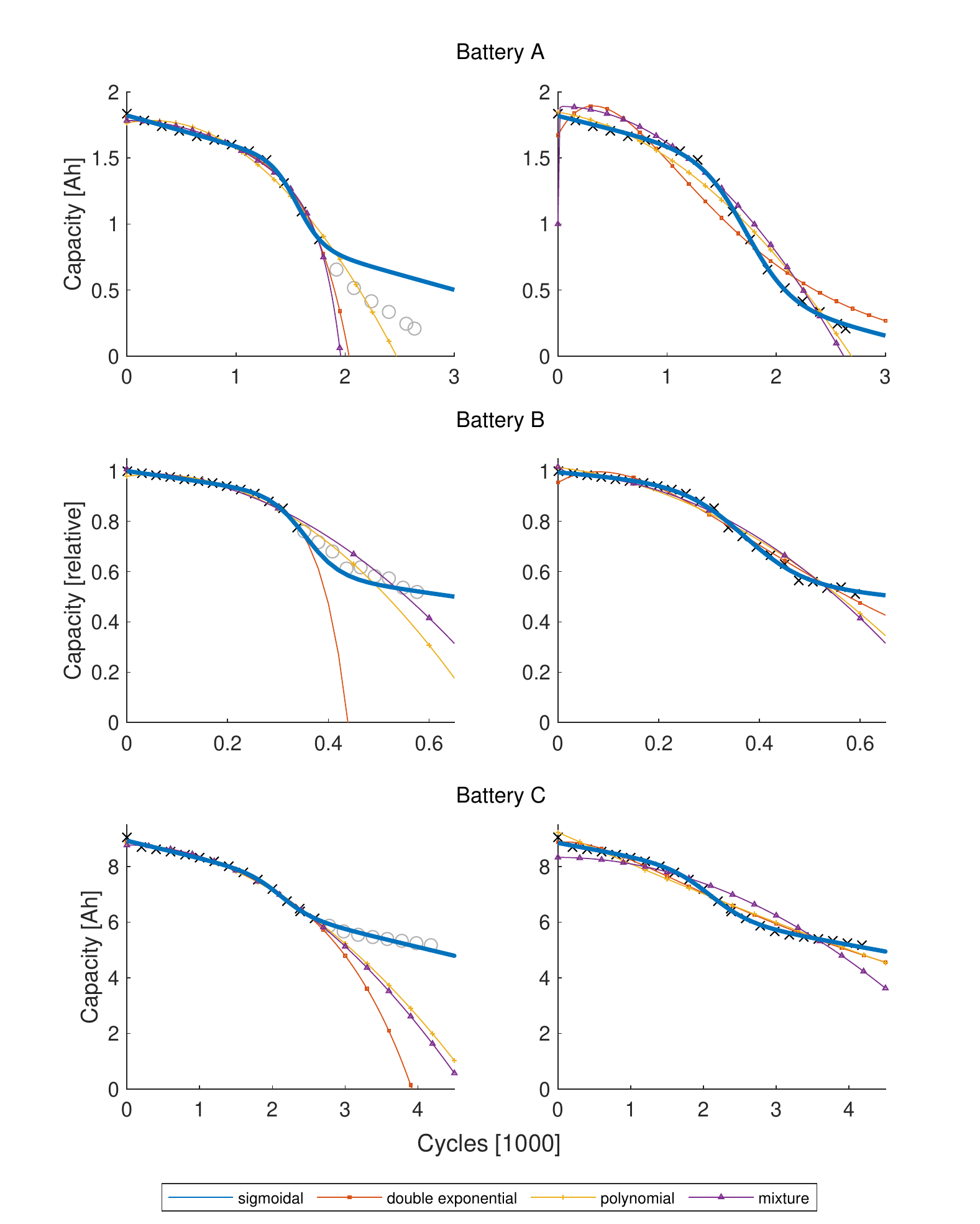}
	\caption{Fits of every discussed model on battery A from \cite{baumhoefer:2014}, battery B from \cite{harrisharris:2017}, and battery C from \cite{lewerenz:2017}. Here, data points up to the first bend (left; A: 12 (data points), B: 350, C: 15) or all data points (right; A: 18, B: 592, C: 23) were used for the fitting; unused data points are marked with a `$\circ$' instead of a `$\times$' on the left. For battery B, there is capacity information available for every cycle and only a fraction of those points is plotted to preserve readability.}
	\label{fig:comp_model_fits}
\end{figure}

As mentioned in Section \ref{s:sigmoid}, all known models, i.e., \eqref{eq:1}, \eqref{eq:2}, and \eqref{eq:3}, are suitable in situations where the capacity degradation of a battery shows only one bend. In Figure \ref{fig:comp_model_fits}, we compare the resulting fits of the three mentioned models to the fitted sigmoidal model for the batteries A, B, and C. The respective optimal parameter vector is generated by an ordinary linear least squares approach for the polynomial model, by the methodology described in the previous section for the sigmoidal model, and in the same manner in an adapted form for the double exponential and the mixture model.
In the left column, only data points up to and including the first bend were used for the fitting (the ignored data points are marked with a `$\circ$') whereas in the right column, all measured data points were included. The parameters of the fitted sigmoidal model for batteries A and B can be found in Table \ref{t:fitted_parameters}.

\begin{figure}[h]
	\centering
	\includegraphics[scale=0.85]{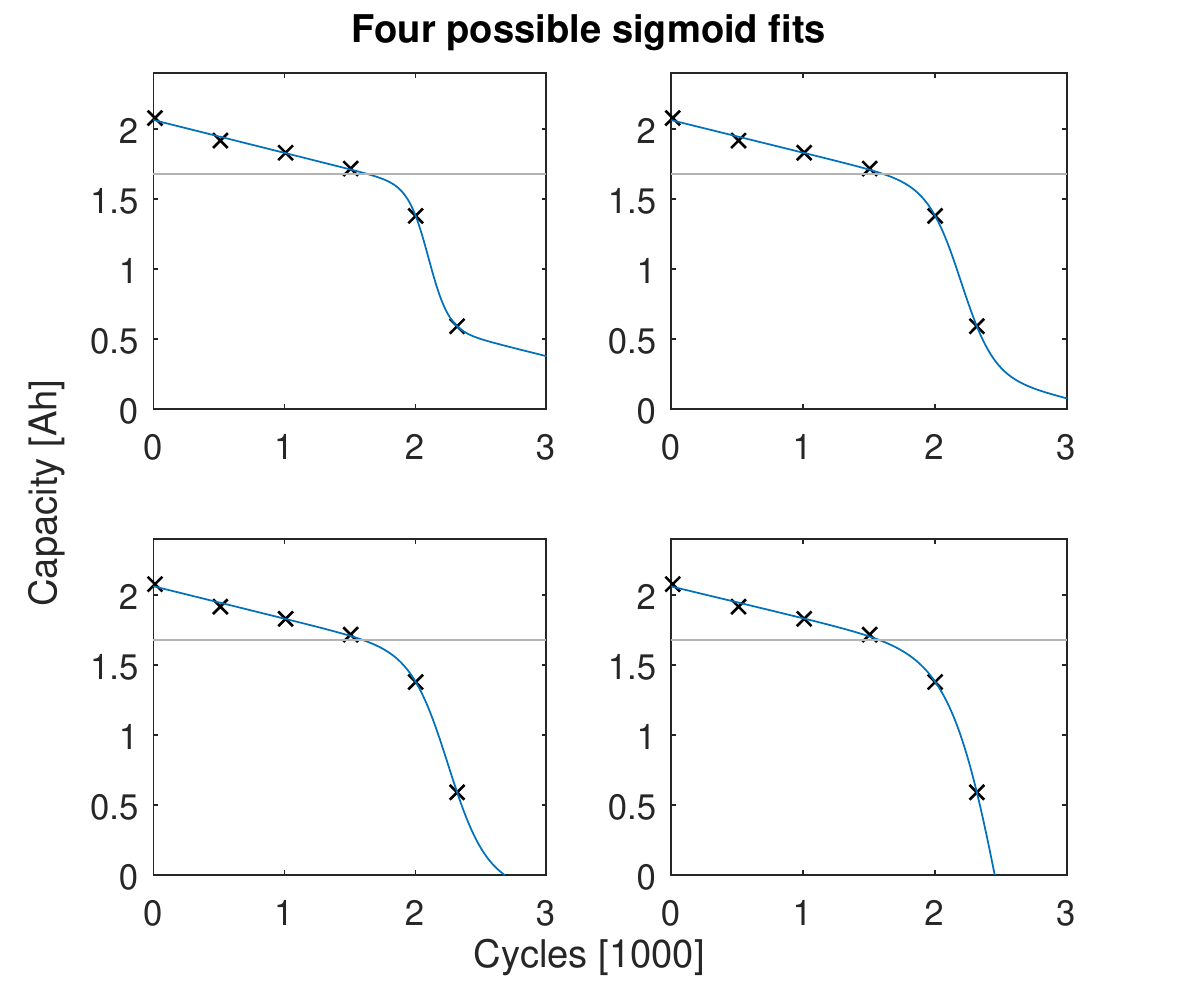}
	\caption{Four possible sigmoid fits on data from a battery from \cite{schmalstieg:2014}, cycled between 50\% and 100\% SOC.}
	\label{fig:4_sigmoid_schmalstieg}
\end{figure}

We can observe that, in the former situation, all models provide appropriate fits which are almost indistinguishable in the cycle range of used measurements. However, the right column displays that only the sigmoidal model is able to fit data indicating a second bend in the capacity degradation. All other models show serious deviations from the observed degradation path, especially when extrapolating to the right. This may lead to a severe underestimation of the future performance of the respective battery, which could be an issue, e.g., in the context of second life use of batteries. Hence, the sigmoidal model is more flexible in the sense that it can depict the capacity degradation in a usual test setup stopping at a remaining capacity of 70-80\% as well as in a situation where a battery is observed until a low level of remaining capacity. The high flexibility of the sigmoidal model is also illustrated in Figure \ref{fig:4_sigmoid_schmalstieg} which presents several possible fits, all applied to the same six data points, differing in the behavior after the last measurement. All fits produce nearly identical sums of squared errors (between 0.0011 and 0.0014) indicating a satisfying fit in each case. This means that the sigmoidal model covers a wide range of possible further developments of the curve equally well, including the case where the capacity drops down to 0 without showing an upward bend; see Figure \ref{fig:4_sigmoid_schmalstieg}, bottom right.

In Table \ref{t:fitted_parameters}, parameters of the fitted sigmoidal model for a single battery and for all batteries of the data sets from \cite{baumhoefer:2014} and \cite{harrisharris:2017} are compared in different censoring situations where only measurements over a threshold of e.g.\ 80\% of the initial capacity are considered. This corresponds to an experiment which is terminated after the respective remaining capacity level is reached. The fit for the parameter $\beta_1$ is seen to remain nearly constant despite right censoring of the data, because it corresponds to the initial capacity (cf.\ Figure \ref{fig:param_sigmoid}), i.e., 1.82 for the absolute capacity measurements (in Ah) from \cite{baumhoefer:2014} and 1.00 for the relative capacity measurements w.r.t.\ the inital capacity from \cite{harrisharris:2017}. On the contrary, $\hat{\beta}_3$ changes significantly for different censoring thresholds. More precisely, if the change of curvature is not observed due to censoring, the sigmoidal model reduces the impact of the logistic shape resulting in values of $\hat{\beta}_3$ close to zero and larger values of $\hat{\beta}_2$.

\begin{table}[h]
	\centering
	\begin{tabular}{*{8}{c}}
		\toprule
		Data set & 
		Battery    & Censoring & $\hat{\beta}_1$ & $\hat{\beta}_2$ & $\hat{\beta}_3$ & $\hat{\beta}_4$ & $\hat{\beta}_5$\\\otoprule
		\multirow{6}{*}{Baumhöfer et al.\ \cite{baumhoefer:2014}} & \multirow{3}{*}{A} & none & 1.82 & 0.20 & 1.06 & 1.72 & 0.21\\
		&& from 50\% & 1.82 & 0.24 & 0.41 & 1.46 & 0.08\\
		&& from 80\% & 1.82 & 0.24 & 0.00 & 3.64 & 0.06\\
		
		\cmidrule{2-8}
		&\multirow{3}{*}{all} & none & 1.82 & 0.21 & 1.03 & 1.57 & 0.26\\
		&& from 50\% & 1.83 & 0.28 & 0.36 & 1.29 & 0.09\\
		&& from 80\% & 1.83 & 0.26 & 0.02 & 0.94 & 0.00\\
		
		\midrule
		\multirow{6}{*}{Harris et al.\ \cite{harrisharris:2017}} & \multirow{3}{*}{B} & none & 1.00 & 0.16 & 0.39 & 0.37 & 0.06\\
		&& from 70\% & 1.00 & 0.30 & 0.19 & 0.33 & 0.02\\
		&& from 80\% & 1.00 & 0.27 & 0.18 & 0.33 & 0.03\\
		
		\cmidrule{2-8}
		&\multirow{3}{*}{all} & none & 1.00 & 0.00 & 0.63 & 0.53 & 0.18\\
		&& from 70\% & 1.00 & 0.31 & 0.04 & 0.30 & 0.02\\
		&& from 80\% & 1.00 & 0.24 & 0.03 & 0.25 & 0.04\\
		\bottomrule
	\end{tabular}
	\caption{Exemplary fits $\hat{\beta}_k$, $k \in\{1,\dots,5\}$ of the parameters in the sigmoidal model, as described in Section \ref{s:model_est} for battery A and all batteries from \cite{baumhoefer:2014} as well as for battery B and all batteries from \cite{harrisharris:2017} in different censoring situations: all data points used, only data until 50\%/70\% remaining capacity used, only data until 80\% remaining capacity used.}
	\label{t:fitted_parameters}
\end{table}

\subsection{Lifetime prediction and confidence intervals}

\begin{figure}[h]
	\begin{subfigure}{.5\textwidth}
		\centering
		\includegraphics[width=\textwidth]{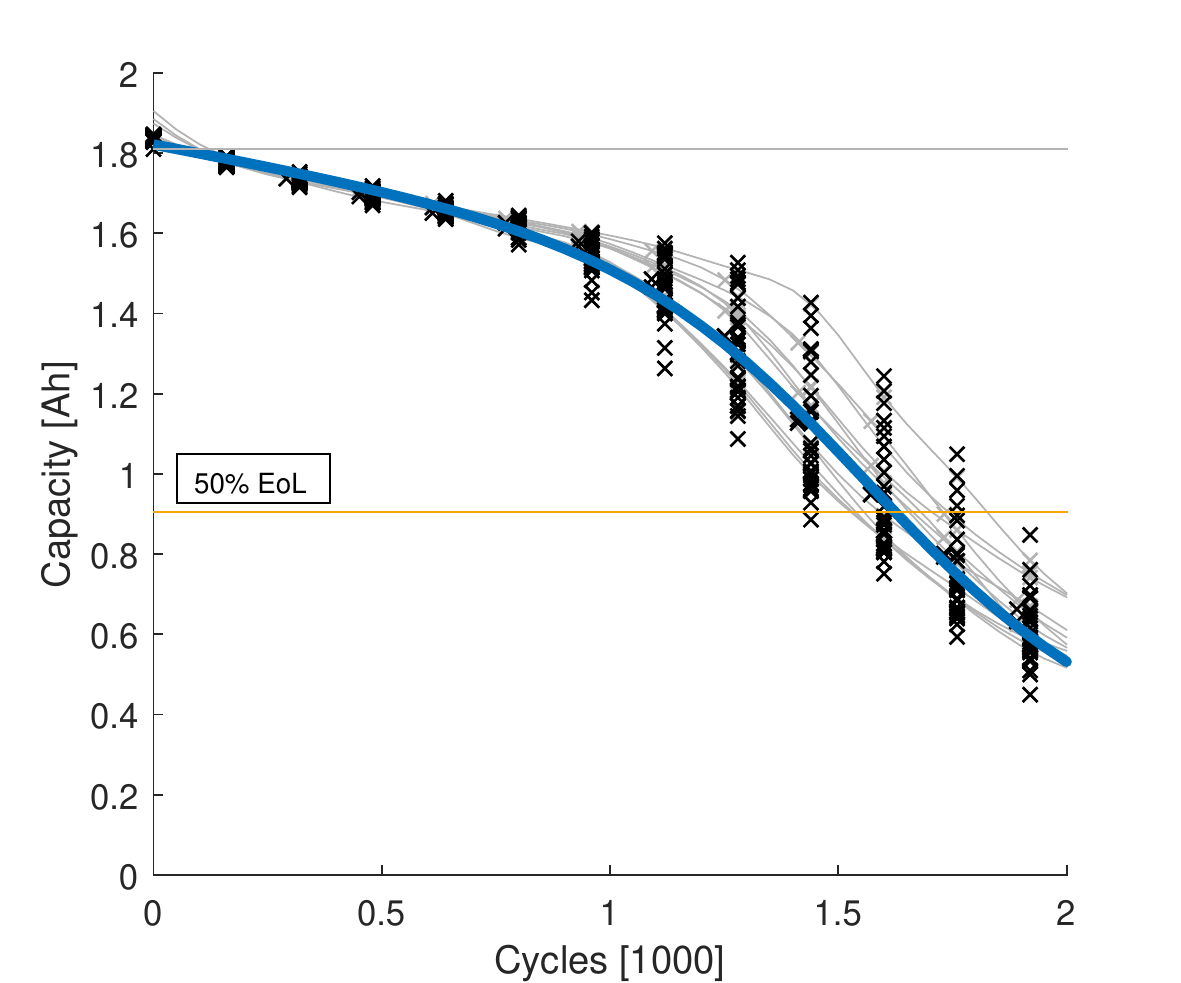}
	\end{subfigure}
	\begin{subfigure}{.5\textwidth}
		\centering
		\includegraphics[width=\textwidth]{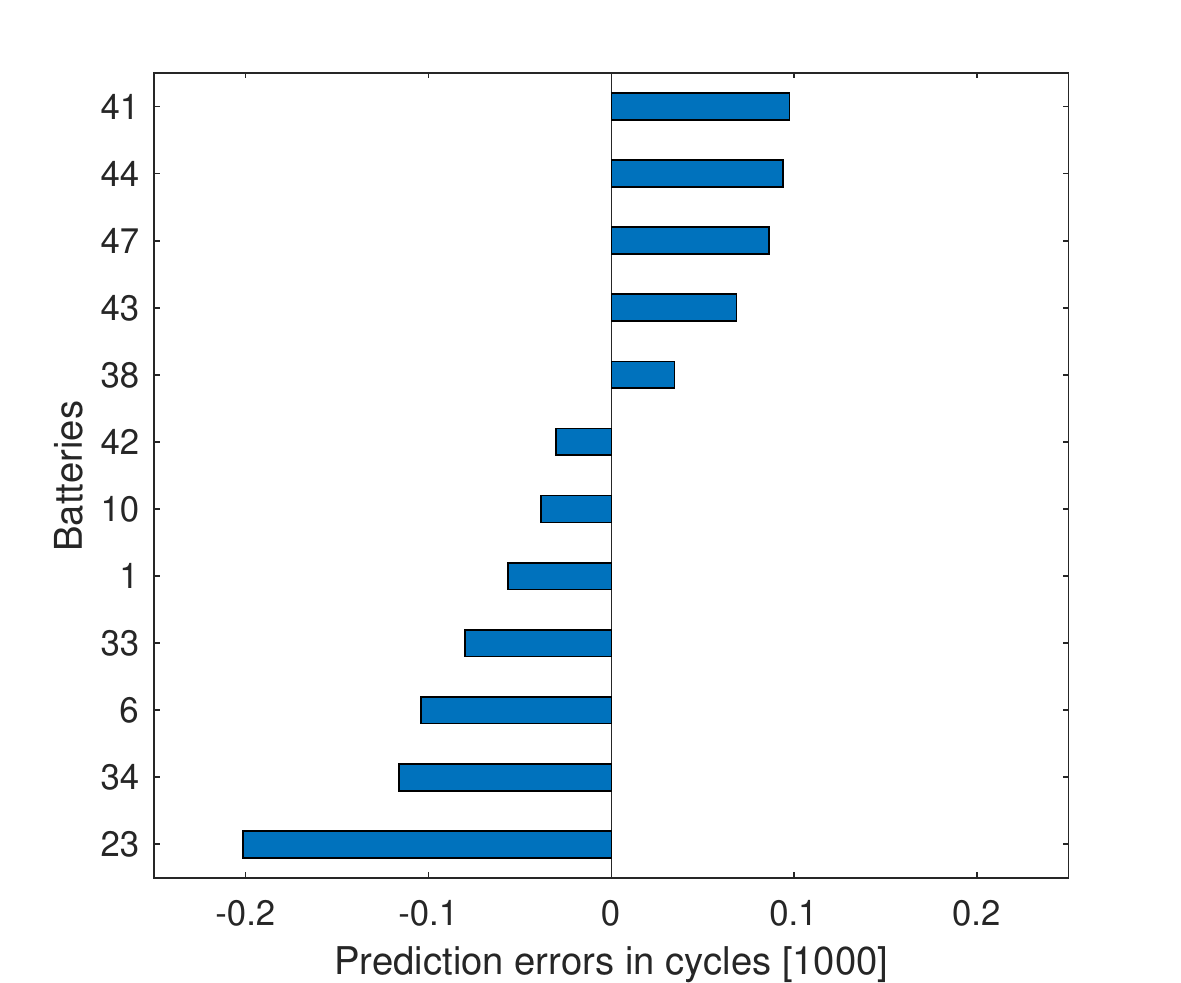}
	\end{subfigure}	
	\caption{Using 36 batteries from \cite{baumhoefer:2014} to predict 50\%-EoL (left) and corresponding prediction errors of the remaining 12 batteries (right).}
	\label{fig:estim_errors}
\end{figure}

\begin{figure}[h]
	\begin{subfigure}{.5\textwidth}
		\centering
		\includegraphics[width=\textwidth]{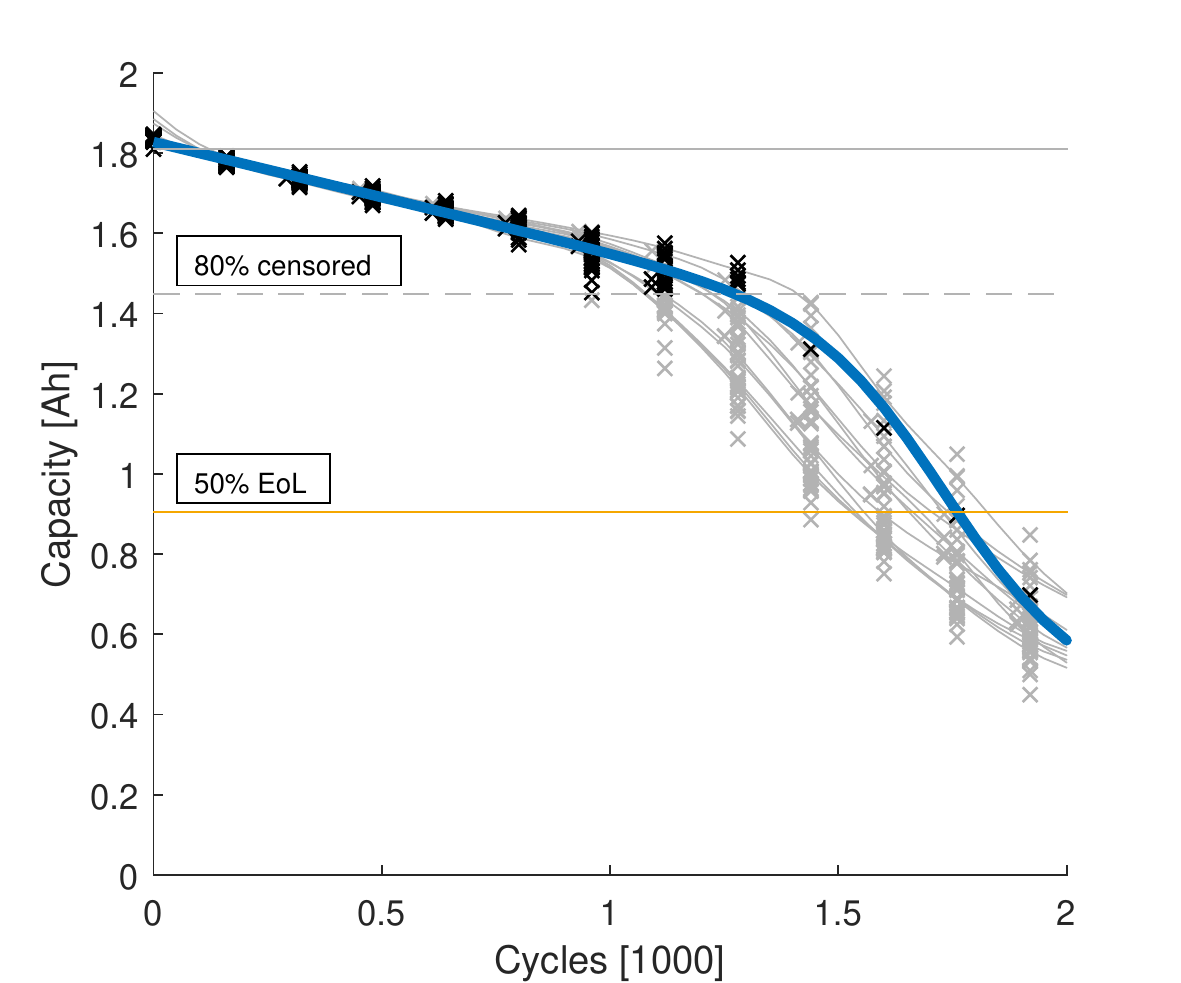}
	\end{subfigure}
	\begin{subfigure}{.5\textwidth}
		\centering
		\includegraphics[width=\textwidth]{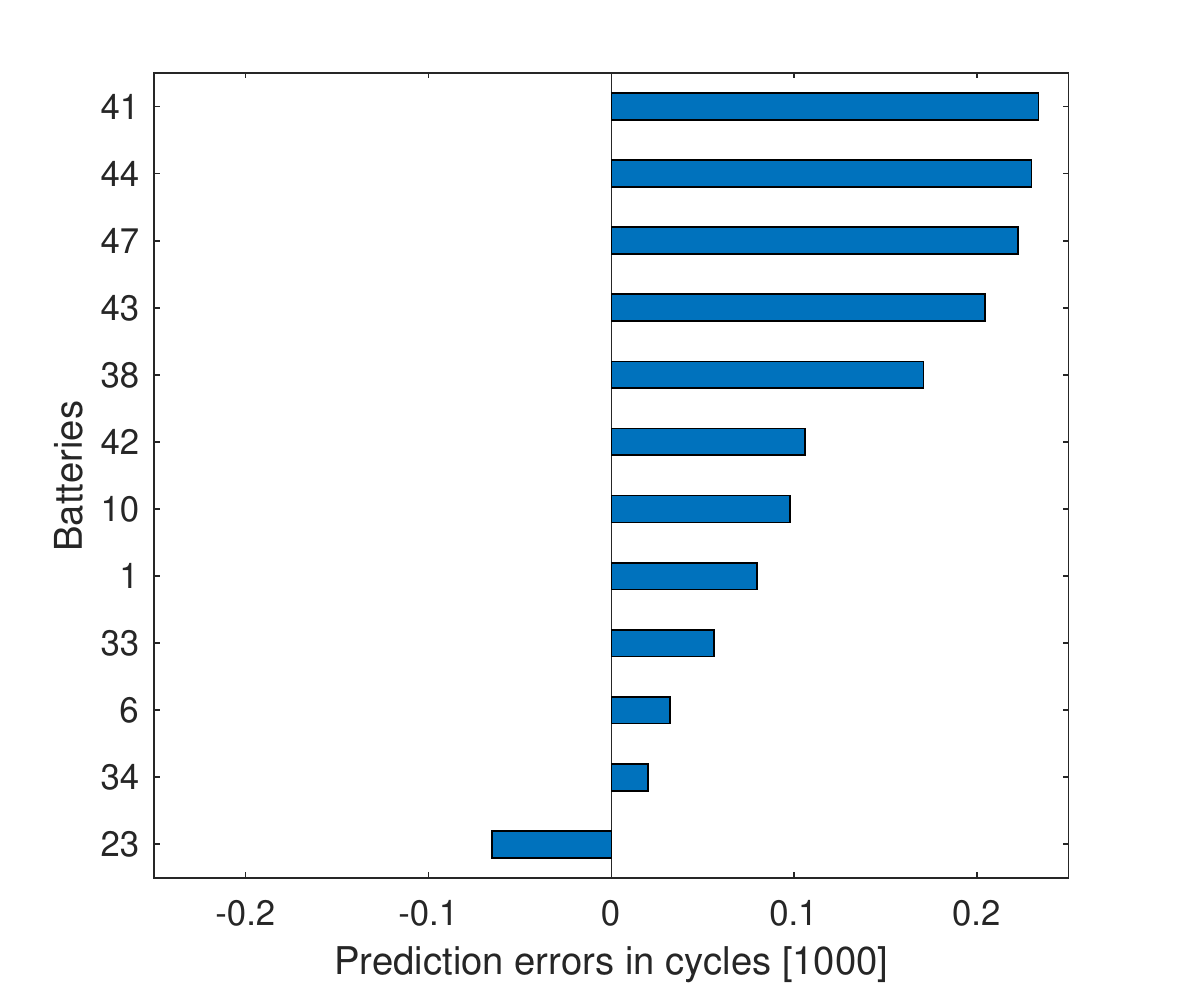}
	\end{subfigure}	
	\caption{Using 36 batteries from \cite{baumhoefer:2014} to predict 50\%-EoL (left) and corresponding prediction errors of the remaining 12 batteries (right). Here, the data in the training sample from all but one battery are censored after $80\%$ of the initial capacity.}
	\label{fig:estim_errors_cens}
\end{figure}

In the following, we give an exemplary application of the lifetime prediction method described in Section \ref{ss:estimation_prediction}. To evaluate the prediction accuracy by cross-validation, we take a subset $T\subsetneq\{1,\dots,48\}$ of the batteries and treat the corresponding index subset $S=\{j\in\{1,\dots,n\} \mid (x_j,y_j) \text{ belong to a battery in } T \}$ of the data from \cite{baumhoefer:2014} as training sample and the complement $S^c=\{1,\dots,n\}\setminus S$ as testing sample. Therefore, we perform a combined sigmoidal fit for all batteries in \(T\), i.e., we calculate
\begin{equation*}
\hbbeta_{S} = \argmin_{\bbeta\in(0,\infty)^5}\sum_{j \in S}\big(f(x_j; \bbeta)-y_j\big)^2
\end{equation*}
as described in Section \ref{ss:estimation_prediction}. The procedure for a specific choice of 36 batteries, i.e., \(\vert T \vert = 36\), is demonstrated in Figure \ref{fig:estim_errors}. The prediction errors \(\hx_{0.5} - x_{0.5,i}\), \(i \in T^{c}\), for $50\%$ EoL are shown on the right. The exact failure time \(x_{0.5,i}\) is derived from a spline interpolation of all data points belonging to the given battery. If we apply censoring to the battery data and apply the same prediction approach, the estimation of the attainment of any capacity level below the censoring threshold shows similar instabilities like the extrapolation for a single battery. However, if we include uncensored data of just one battery in the training sample as shown in Figure \ref{fig:estim_errors_cens}, the prediction method provides reasonable results which are only slightly worse than in the fully uncensored case. This is also observed in Table \ref{t:prediction_acu}, which gives the mean squared error (MSE), the root mean squared error (RMSE), the mean error (ME), and the mean absolute error (MAE) for different EoL thresholds \(q \in (0,1)\), where for any set of batteries $T\subsetneq\{1,\dots,48\}$
\begin{align}\label{eq:mseetc}
\begin{split}
\text{MSE} &= \frac{1}{\vert T^c \vert}\sum_{i \in T^c}\big(\hx_q-x_{q,i}\big)^2,\\
\text{RMSE} &= \sqrt{\frac{1}{\vert T^c \vert}\sum_{i \in T^c}\big(\hx_q-x_{q,i}\big)^2},\\
\text{ME} &= \frac{1}{\vert T^c \vert}\sum_{i \in T^c}\big(\hx_q-x_{q,i}\big),\\
\text{MAE} &= \frac{1}{\vert T^c \vert}\sum_{i \in T^c}\big\vert \hx_q-x_{q,i} \big\vert
\end{split}
\end{align}
for \(\hx_q = f^{-1}(y_q;S)\), where \(S\) is the corresponding index subset to \(T\). Here, \(\hx_q\) and \(x_{q,i}\) are given in 1000 cycles.

For Table \ref{t:prediction_acu}, 100 sets of batteries \(T\) are randomly selected and the named quantities are calculated each time for both scenarios (24/36 batteries) described above. The values given in Table \ref{t:prediction_acu} are the means of the 100 respectively resulting  outcomes. Note that the prediction accuracy is not much affected by the size of the training sample in both censoring situations.

As depicted in Figure \ref{fig:harris_daten}, the data set from \cite{harrisharris:2017} does not show a sigmoidal shape in the capacity degradation for the vast majority of the tested batteries. If we apply the same prediction approach as seen before, using the sigmoidal model and the double exponential model from \eqref{eq:1} in comparison as basis, the results, presented in Table \ref{t:prediction_acu_harris_comp}, are very similar. Hence, the sigmoidal model and the corresponding prediction methods are also applicable in data situations where no striking sigmoidal shape is observed. Censoring scenarios were not considered for the data set from \cite{harrisharris:2017}, because it only consists of 24 batteries partially showing a final remaining capacity over 80\% and censoring would lead to an insufficient number of observed failures for cross-validation.\medskip   
 
\begin{figure}[h]
	\begin{subfigure}{.5\textwidth}
		\centering
		\includegraphics[width=\textwidth]{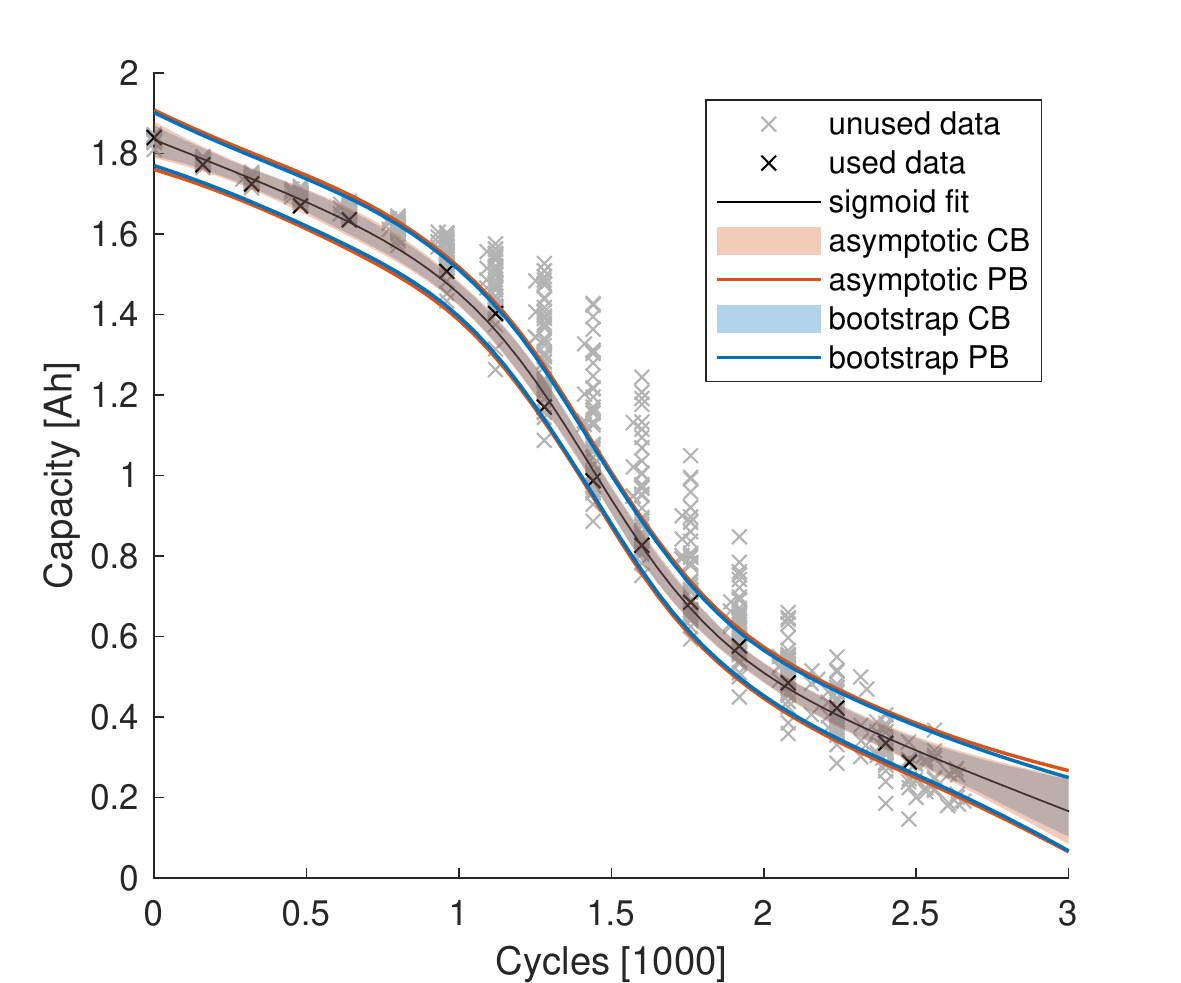}
	\end{subfigure}
	\begin{subfigure}{.5\textwidth}
		\centering
		\includegraphics[width=\textwidth]{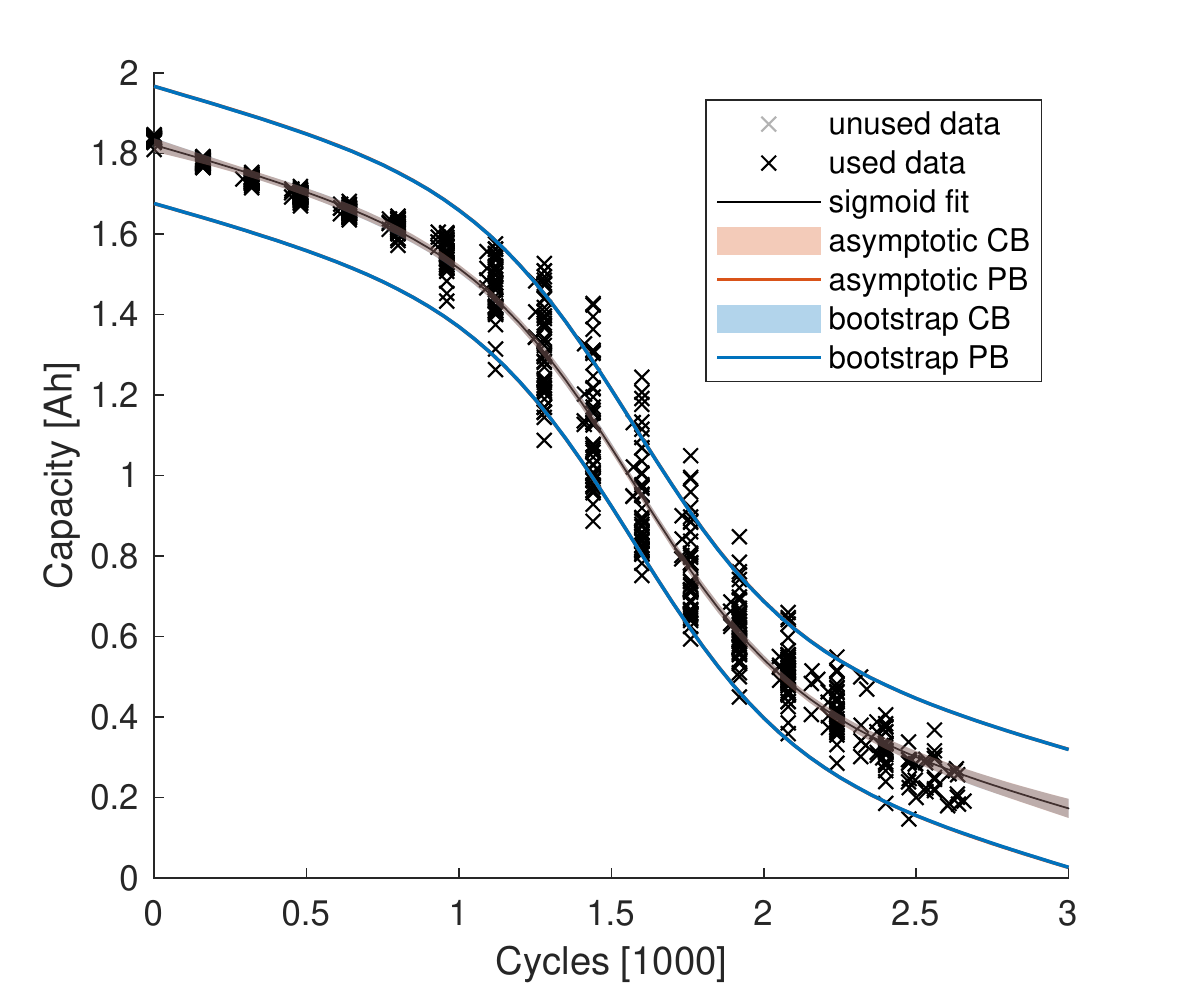}
	\end{subfigure}
	\caption{Pointwise bootstrap and asymptotic confidence bands (CBs) and prediction bands (PBs) for the data from \cite{baumhoefer:2014}, where only the data from battery A (left) or all data points (right) are used.}
	\label{fig:confidence_bands}
\end{figure}

\begin{figure}[h]
	\begin{subfigure}{.5\textwidth}
		\centering
		\includegraphics[width=\textwidth]{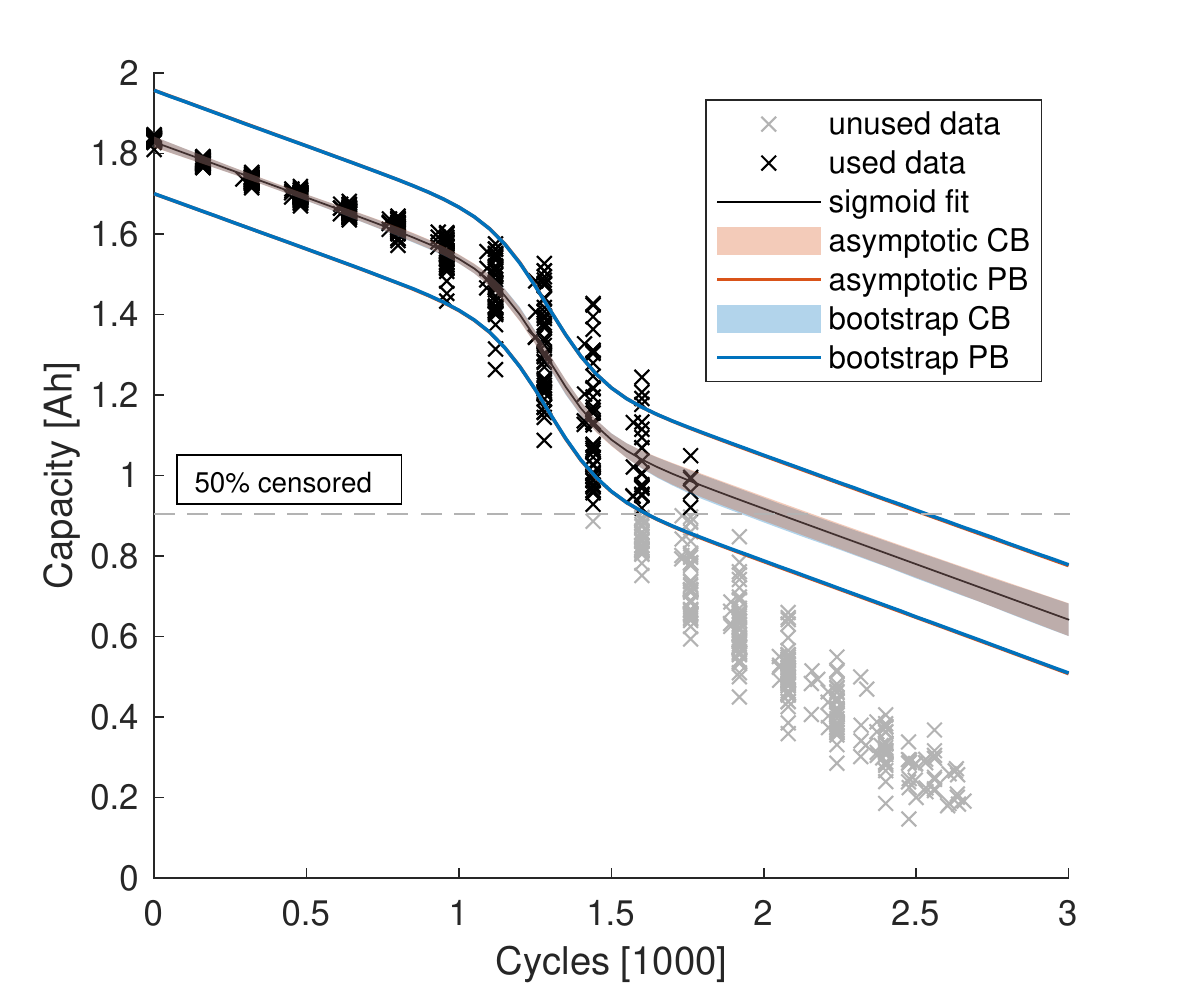}
	\end{subfigure}
	\begin{subfigure}{.5\textwidth}
		\centering
		\includegraphics[width=\textwidth]{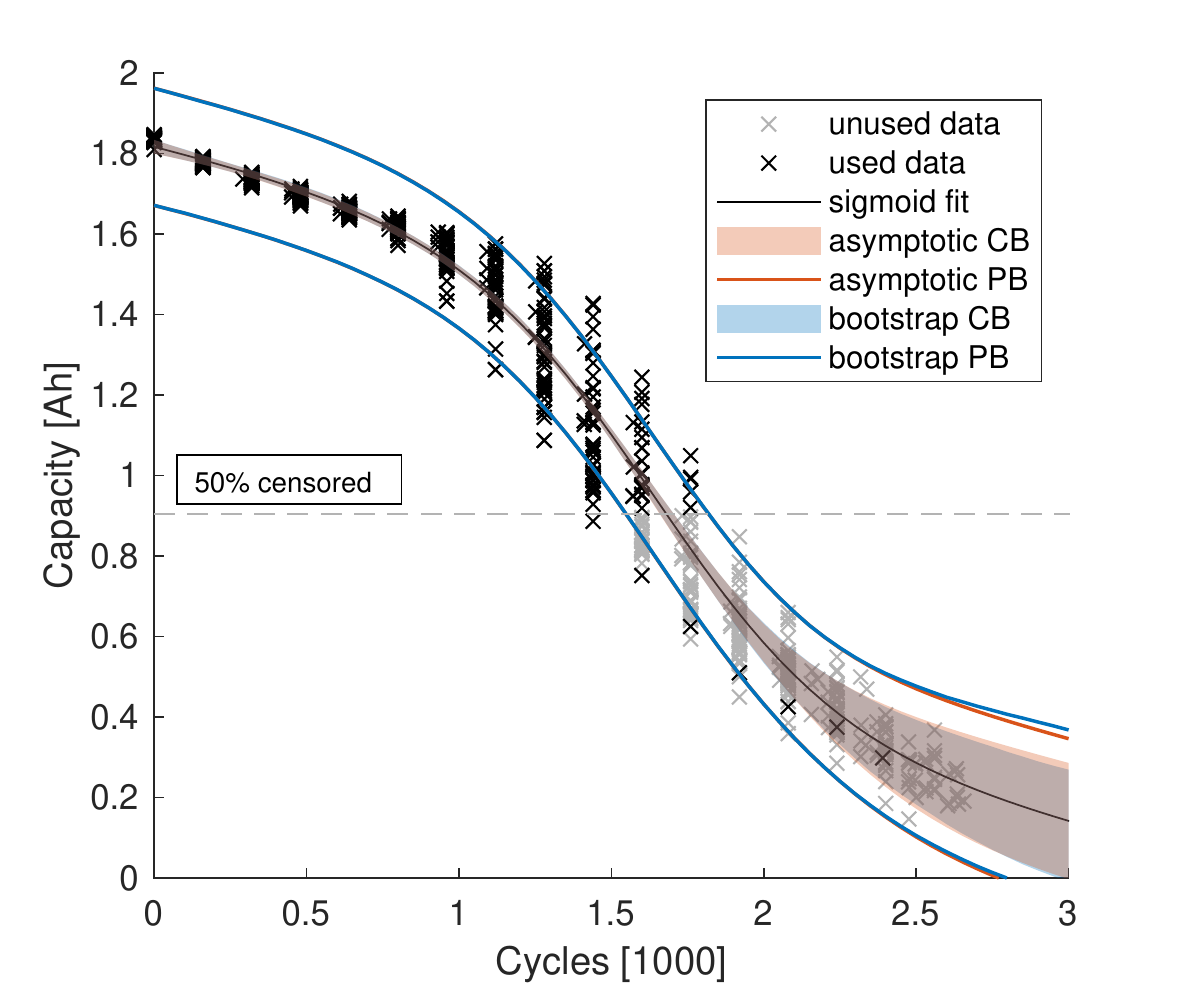}
	\end{subfigure}
	\caption{Pointwise bootstrap and asymptotic confidence bands (CBs) and prediction bands (PBs) for the data from \cite{baumhoefer:2014} based on data from all batteries, censored after 50\% capacity (left) and, additionally, uncensored data from one battery (right).}
	\label{fig:confidence_bands_censoring}
\end{figure}

In Figure \ref{fig:confidence_bands}, pointwise confidence and prediction intervals from Sections \ref{ss:confidence_intervals_asymptotic} and \ref{ss:confidence_intervals_bootstrap}, referred as confidence and prediction bands (CB and PB),  are presented for the data from battery A and all batteries from \cite{baumhoefer:2014}. Note that the respective asymptotic and bootstrap bands are nearly identical. Here, the confidence intervals become shorter whereas the prediction intervals become longer when data from all batteries are used instead of from a single battery, only. Since the confidence band estimates the expected capacity degradation, more information is available when including more data points. On the other hand, different batteries show different capacity degradations (cf.\ Figure \ref{fig:baumhoefer_cap_2}) and therefore, measurement errors become larger when including data from different batteries, resulting in wider prediction bands. In each case, the confidence bands are very narrow, meaning that the regression function describing the expected capacity degradation can be estimated well from the available data. In Figure \ref{fig:confidence_bands_censoring}, a similar analysis is done with data from all batteries from \cite{baumhoefer:2014}, but with censoring after 50\% of the initial capacity on the left, and with additional complete data from a single battery on the right. Again, asymptotic and bootstrap procedures yield nearly identical intervals. In analogy to the observation for lifetime prediction, interval estimation of the capacity degradation fails in the first case, but yields reasonable results in the latter case.

\begin{table}[h]
	\centering
	\begin{tabular}{*{12}{c}}
		\toprule
		\multirow{2}{*}{EoL}	& \multirow{2}{*}{sample size} & \hspace{2mm}	& \multicolumn{4}{c}{full data} & \multirow{2}{*}{\hspace{3mm}censoring\hspace{3mm}} & \multicolumn{4}{c}{censored data (one complete)}\\
		\cmidrule(lr){4-7}
		\cmidrule(lr){9-12}
		& && MSE 	& RMSE 	& ME	& MAE 	&& MSE 	& RMSE 	& ME	& MAE 	\\
		\otoprule
		\multirow{2}{*}{85\%} 	& 75\% 			&& .012	& .109 	& .026	& .096 	& \multirow{2}{*}{90\%} & .022	& .141	& .005	& .118 \\
		& 50\% 			&& .012	& .111	& .019	& .096	&& .023	& .146 	& .013	& .122 \\
		\midrule
		\multirow{2}{*}{80\%} 	& 75\% 			&& .012	& .109 	&.020	& .095	& \multirow{2}{*}{90\%} & .020	&  .137	& $-.020$	& .112 \\
		& 50\% 			&& .013	& .112	& .026	& .097	&& .026	& .153 	& .019	& .128 \\
		\midrule
		\multirow{2}{*}{50\%} 	& 75\% 			&& .013	& .111	& .031	& .098	& \multirow{2}{*}{75\%} & .020	& .133 	& .043	& .114 \\
		& 50\% 			&& .013	& .112	& .025	& .097	&& .022	& .140 	& .055	& .121 \\
		\midrule
		\multirow{2}{*}{40\%} 	& 75\% 			&& .012	& .108	&.023	& .093	& \multirow{2}{*}{70\%} & .022	& .141 	& .077	& .124 \\
		& 50\% 			&& .012	& .111	& .020	& .096	&& .022	& .140 	& .072	& .123 \\
		\bottomrule
	\end{tabular}
	\caption{Results from a cross-validation for the accuracy of lifetime prediction based on the sigmoidal model, for several EoL thresholds and different sizes of the training sample based on the 48 batteries from \cite{baumhoefer:2014}. Presented are the mean squared error (MSE), root mean squared error (RMSE), mean error (ME), and mean absolute error (MAE) (cf.\ \eqref{eq:mseetc}), averaged over 100 random training samples. Prediction is based on fully observed data as well as on data censored after certain percentages of the initial capacity, where only one battery is fully observed. Here, batteries not observed until the respective EoL threshold were excluded from the analysis.}
	\label{t:prediction_acu}
\end{table}

\begin{table}[h]
	\centering
	\begin{tabular}{*{12}{c}}
		\toprule
		\multirow{2}{*}{EoL}	& \multirow{2}{*}{sample size} & \hspace{2mm}	& \multicolumn{4}{c}{sigmoid} & \hspace{3mm} & \multicolumn{4}{c}{double exponential}\\
		\cmidrule(lr){4-7}
		\cmidrule(lr){9-12}
		& && MSE 	& RMSE 	& ME	& MAE 	&& MSE 	& RMSE 	& ME	& MAE 	\\
		\otoprule
		\multirow{2}{*}{85\%} 	& 75\% 			&& .011	& .104 	& $-.036$	& .091 	&  & .011	&  .104	& $-.035$	&  .091 \\
		& 50\% 			&& .012	& .106	& $-.033$	& .092	&& .012	& .107 	& $-.034$	& .093 \\
		\midrule
		\multirow{2}{*}{80\%} 	&  75\% 			&& .012	& .105 	& $-.036$	& .091	&& .012 & .105	& $-.035$ 	& .092  \\
		& 50\% 			&& .011	& .105	& $-.029$	& .090	&& .011	& .105 	& $-.029$	& .091 \\
		\bottomrule
	\end{tabular}
	\caption{Results from a cross-validation for the accuracy of lifetime prediction for several EoL thresholds and different sizes of the training sample based on the 24 batteries from \cite{harrisharris:2017}, based on the sigmoidal and the double exponential model. Presented are the mean squared error (MSE), root mean squared error (RMSE), mean error (ME), and mean absolute error (MAE) (cf.\ \eqref{eq:mseetc}), averaged over 100 random training samples. Here, batteries not observed until the respective EoL threshold were excluded from the analysis.}
	\label{t:prediction_acu_harris_comp}
\end{table}

\section{Discussion}

For the long term capacity degradation behavior of lithium-ion battery cells under cyclic aging, we propose a parametric regression model with the regression function being a linear combination of a linear and a logistic function.
The sigmoidal model allows for interpretation of all of its five parameters and shows high flexibility in fitting different kinds of capacity degradation paths. In particular, the typical concave shape of short-term experiments showing only one bend and the sigmoidal shape showing a second bend occurring in long-term experiments can both be fitted well.  
The sigmoidal model is similar in performance compared to other established models in the short-term situation but is the only satisfactory model for the log-term capacity behavior.
The regression set-up enables statistical estimation and prediction procedures for battery characteristics of interest. Examples are given for parameter estimation, confidence and prediction intervals for the remaining capacity after a given number of experimental cycles as well as for lifetime prediction under arbitrary EoL thresholds.
In practical situations, where only short-term information is available, the model is able to predict the further development well if at least the degradation path of a single battery is known entirely and included in the training sample, provided it is a typical degradation path. By applying the proposed procedure, criteria are at hand to raise statistical precision, to reduce the number of test objects as well as to shorten the total experimental time.


\end{document}